# Negative refraction and sub-wavelength imaging using transparent metal-dielectric stacks

**Michael Scalora, Giuseppe D'Aguanno, Mark Bloemer**

*Charles M. Bowden Research Center, AMSRD-AMR-WS-ST, Research, Development, and Engineering Center,*

*Redstone Arsenal, AL 35898-5000*

michael.scalora@us.army.mil

**Domenico de Ceglia**

*Dipartimento di Elettrotecnica ed Elettronica, Politecnico di Bari, Via Orabona 4, 70124 Bari, Italy*

**Marco Centini**

*INFM at Dipartimento di Energetica, Universita di Roma 'La Sapienza', Via A. Scarpa 16, 00161 Roma, Italy*

**Nadia Mattiucci, Neset Akozbek**

*Time Domain Corporation, Cummings Research Park 7057 Old Madison Pike Huntsville, Alabama 35806, USA and*

*Charles M. Bowden Research Center, AMSRD-AMR-WS-ST, Research, Development, and Engineering Center,*

*Redstone Arsenal, AL 35898-5000*

**Mirko Cappeddu**

*Universita' degli Studi di Catania, Dipartimento di Ingegneria Elettrica Elettronica e dei Sistemi, Via A. Doria 6, 95125 Catania - Italy*

**Joseph W. Haus**

*Electro-Optics Program, University of Dayton, Dayton, OH 45469-0245, USA*

**Abstract:** Negative refraction is known to occur in materials that simultaneously possess a negative electric permittivity and magnetic permeability; hence they are termed negative index materials. However, there are no known natural materials that exhibit a negative index of refraction. In large part, interest in these materials is due to speculation that they could be used as perfect lenses with superresolution. We propose a new way of achieving negative refraction with currently available technology, based on transparent, metallo-dielectric multilayer structures. The advantage of these structures is that both tunability and transmission (well above 50%) can be achieved in the visible wavelength regime. We demonstrate both negative refraction and sub-wavelength resolution in these structures. Our findings point to a simpler way to fabricate a material that exhibits negative refraction. This opens up an entirely new path not only for negative refraction, but also to expand the exploration of wave propagation effects in metals.

**Introduction**

Historically, the propagation of light in metals has been considered only to some extent because bulk metals are highly opaque at almost all wavelengths across the entire electromagnetic spectrum. A quasi-transparent region naturally occurs near the plasma frequency, where the real part of the index of refraction approaches unity, and absorption is minimized [1]. Understanding how light propagates through metals has taken on a special significance since the prediction that an object, illuminated with TM-polarized light, may form a super-resolved image of itself on the other side of a thin metal layer [2]. The image is formed by "negative refraction" [3] that seemingly occurs inside the metal, as discussed in reference [2]. The metal focuses a wave-front to a tight spot, thus exciting large spectral components (k-space), including evanescent modes. The argument made is that the presence of evanescent modes at the focus, in addition to the propagating modes, makes it possible for the image to be resolved beyond the limits imposed by ordinary optical materials and by diffraction [2].

The basic mechanism described in reference [2] is based on a change in direction of the longitudinal component of the electric field inside the metal, which fosters a change in direction of the Poynting vector. In other words, the Poynting vector of a TM-polarized beam, incident at an angle on a metal surface, or layer, appears to *refract* with a negative angle inside the metal. It is the continuity of the longitudinal component of the electric displacement field, **D**=ε**E,** at the boundary that forces the change in the direction of propagation of the Poynting vector. Then, if the metal layer is sufficiently thick, and the object is sufficiently close, the light may form a first focus inside the metal layer, and a second focus outside it, where the object is imaged with features that exceed the diffraction limit, i.e. superresolution. The effect predicted in reference [2] has been experimentally observed and verified [4, 5].

A shortcoming with the single-layer super-lens is the opacity of metals. Therefore, the first theoretical prediction was made, and experiments carried out, using silver layers 40-50nm thick, below but near the plasma frequency, which occurs at ~320nm, where the dielectric constant of silver is negative, and the material is still somewhat transparent. The modus operandi of the transmission process is based on the excitation of surface modes, which evanescently couple the light to the other side of the barrier. In general, transmittance through a single metal layer is rather low, and gets rapidly worse in the visible range and beyond as the wavelength increases. The opacity of metals has thus driven researchers to seek negative refraction and super-lensing in a roundabout way, by first engineering a negative index material in the microwave regime using metallic rods and loops to induce an electric and magnetic resonance [6-8], and then by seeking a way to scale down the size of magnetic circuits using nano-wires or nano-strips, for example, in order to access the much coveted visible region [9]. However, the containment of material and scattering losses, along with the successful achievement of size reduction, remain formidable challenges.

In this article we report that there is a relatively easy way to obtain negative refraction and sub-wavelength resolution for TM-polarized light throughout the visible range, using transparent, metallo-dielectric stacks [10-14]. These are multilayer structures that may contain several metallic layers, where the thickness of each layer may be tens of nanometers, and may even be a good fraction of the incident wavelength (in some cases, *each* metal layer may be 50-60nm thick, thus easily approaching 10-15% of the size of the wavelength), so that total metal thickness may easily reach into the hundreds of nanometers. Remarkably, the structure can be designed so that it remains transparent within pre-selected, broad wavelength ranges in the visible region and beyond. The location of the transparency window usually depends on the thickness of the intervening dielectric material, and can be engineered rather easily by using a combination of appropriate metals, and/or dielectric materials, from the visible well into the mid-infrared range. For instance, one may use almost any dielectric or semiconductor material, in combination with aluminum (for work in the ultraviolet range), silver (best used in the visible part of the spectrum), gold, and copper (both better suited for near-infrared applications). As an example, an 11-layer, $Ag/MgF_2$ stack containing 150nm of silver was fabricated [11,12] using thermal evaporation techniques, that featured a peak transmittance of about 52% at 532nm, with a full width at half maximum well in excess of 100nm. Metal layer thicknesses ranged from a low of 20 to a high of 40nm, and so it is possible to further increase transmittance by thinning out Ag layers, and by choosing a material with a relatively high index of refraction, such as ZnO [13] or $TiO_2$. We note that nonlinear effects have also been studied in $Cu/SiO_2$ stacks, similar to those that we describe, both theoretically [15] and experimentally [16]. In reference [16], for example, the localization of the light inside the Cu layers was confirmed by directly measuring nonlinear phase and transmittance changes as a function of incident intensity.

Also recently, an approach that considers metallo-dielectric multilayers for superresolution has been proposed [17]. In that study, metal and dielectric layers were assumed to be infinitesimally thin, and the stack was described as an effective medium. As such, field localization and interference effects of any sort are precluded inside the stack, a fact that stands in contrast to the transparent metal case that we discuss here. In fact, we rely on interference effects and field confinement that occurs inside the stack, within *both* dielectric and metal layers, to achieve negative refraction and sub-wavelength resolution. There are more differences that can be pointed out. For instance, the effective medium approach predicts a strong, anisotropic response of the multilayer stack, in a way that the longitudinal component of the dielectric constant becomes infinite, and the transverse component goes to zero. In this limit, the stack will effectively guide all incident light emanating from the entry surface straight down to the exit side, forming a super-resolved image of the object, as shown in reference [17]. However, the strong anisotropy of the stack also causes the transmitted portion of any beam incident from the outside, at any angle, to refract at zero-angle due to the large, longitudinal effective index of refraction. No negative refraction takes place. In contrast, in our case we find that the refraction angle can be controlled with an appropriate choice of intervening dielectric material and/or metal thickness, and we argue that the effective medium approach is not applicable. As we noted above, the thickness of each layer may easily be 10-15% of the size of the wavelength, and metallic components may account for more than half, sometimes three quarters of the entire structure's length. Under these circumstances, it is difficult to imagine any description of the structures that we propose as an anisotropic, effective medium in the sense of reference [17].

We thus reasonably conclude that while geometry may ruse the eye, our approach and that described in reference [17] are in fact quite different. Here, we discuss effects due to propagation effects that directly coax the light *inside* individually thick metal layers to increase both local energy and momentum values. At the same time, absorption and the evanescent nature of the field are partially suppressed, in favor of a simple resonance tunneling mechanism that takes hold, and that quickly and effectively pushes the light forward. It then becomes nearly impossible to excite surface waves, as the light finds an easy "channel" to the outside. The resonance tunneling mechanism has been experimentally verified in the linear and nonlinear regimes, and they are discussed in references [10-16]. In the sections that follow we will put forward a vector propagation model that accounts for the dispersive nature of the materials in question, including absorption, and show how the process of negative refraction and sub-wavelength resolution occur simultaneously in transparent metal stacks in the visible range.

**The Propagation model**

Although the propagation model is described in details elsewhere [18-20], for completeness we re-propose the most salient points here, as we consider several aspects that may have been treated separately in the references cited. We use a pulse propagation model that integrates the vector Maxwell's equations in the time domain with two spatial coordinates and time. We use the Gaussian system of units, and we assume a TM-polarized incident field of the form:

$$\mathbf{H} = \hat{\mathbf{x}}(\mathcal{H}_x(y,z,t)e^{i(k_z z - k_y y - \omega_0 t)} + c.c)$$
$$\mathbf{E} = \hat{\mathbf{y}}(\mathcal{E}_y(y,z,t)e^{i(k_z z - k_y y - \omega_0 t)} + c.c) + \hat{\mathbf{z}}(\mathcal{E}_z(y,z,t)e^{i(k_z z - k_y y - \omega_0 t)} + c.c)$$
(1)

where $\hat{\mathbf{x}}$, $\hat{\mathbf{y}}$, $\hat{\mathbf{z}}$ are the unit directional vectors; $\mathbf{E}$ and $\mathbf{H}$ are real electric and magnetic fields, respectively; $\mathcal{H}_x(y,z,t)$, $\mathcal{E}_y(y,z,t)$, and $\mathcal{E}_z(y,z,t)$ are general, complex envelope functions; $k_z = |\mathbf{k}|\cos\theta_i$ and $k_y = -|\mathbf{k}|\sin\theta_i$, $|k|=k_0=\omega_0/c$, and $\theta_i$ is the angle of incidence. This choice of carrier wave-vector is consistent with a pulse initially located in vacuum, and we make no other assumptions about the envelope functions. For simplicity, we assume that the medium is isotropic. Following Eqs.(1), the y-component of displacement field $\mathbf{D}$ may be similarly defined as follows: $\mathcal{D}_y = (\mathcal{D}_y(y,z,t)e^{i(k_z z - k_y y - \omega_0 t)} + c.c)$, and may be related to the electric field by expanding the complex dielectric function as a Taylor series in the usual way:

$$\varepsilon(\mathbf{r},\omega) = \varepsilon(\mathbf{r},\omega_0) + \frac{\partial \varepsilon(\mathbf{r},\omega)}{\partial \omega}|_{\omega_0}(\omega - \omega_0) + \frac{1}{2}\frac{\partial^2 \varepsilon(\mathbf{r},\omega)}{\partial \omega^2}|_{\omega_0}(\omega - \omega_0)^2 + ....$$
(2)

Then, for an isotropic medium, a simple constitutive relation may be written as follows:

$$\mathcal{D}_y(\mathbf{r},t) = \int_{-\infty}^{\infty} \varepsilon(\mathbf{r},\omega)\tilde{\mathcal{E}}_y(\mathbf{r},\omega)e^{-i\omega t}d\omega \;;$$
(3)

where $\tilde{\mathcal{E}}_y(\mathbf{r},\omega)$ is the Fourier transform of $\mathcal{E}_y(\mathbf{r},t)$. Assuming that a similar development follows for remaining electric field and the magnetic fields, Maxwell's equations take the following form [18-20]:

$$\gamma \frac{\partial \mathcal{H}_x}{\partial \tau} + i\frac{\gamma'}{4\pi} \frac{\partial^2 \mathcal{H}_x}{\partial \tau^2} - \frac{\gamma''}{24\pi^2} \frac{\partial^3 \mathcal{H}_x}{\partial \tau^3} + ... = i\beta\left(\mu(\tilde{y},\xi)\mathcal{H}_x + \mathcal{E}_z \sin\theta_i + \mathcal{E}_y \cos\theta_i\right) - \frac{\partial \mathcal{E}_z}{\partial \tilde{y}} + \frac{\partial \mathcal{E}_y}{\partial \xi}$$

$$\alpha \frac{\partial \mathcal{E}_y}{\partial \tau} + i\frac{\alpha'}{4\pi} \frac{\partial^2 \mathcal{E}_y}{\partial \tau^2} - \frac{\alpha''}{24\pi^2} \frac{\partial^3 \mathcal{E}_y}{\partial \tau^3} + ... = i\beta\left(\varepsilon(\tilde{y},\xi)\mathcal{E}_y + \mathcal{H}_x \cos\theta_i\right) + \frac{\partial \mathcal{H}_x}{\partial \xi} \quad (4)$$

$$\alpha \frac{\partial \mathcal{E}_z}{\partial \tau} + i\frac{\alpha'}{4\pi} \frac{\partial^2 \mathcal{E}_z}{\partial \tau^2} - \frac{\alpha''}{24\pi^2} \frac{\partial^3 \mathcal{H}_z}{\partial \tau^3} + ... = i\beta\left(\varepsilon(\tilde{y},\xi)\mathcal{E}_z + \mathcal{H}_x \sin\theta_i\right) - \frac{\partial \mathcal{H}_x}{\partial \tilde{y}}.$$

Here, $\alpha = \frac{\partial[\tilde{\omega}\varepsilon(\xi)]}{\partial \tilde{\omega}}\big|_{\omega_0}$, $\gamma = \frac{\partial[\tilde{\omega}\mu(\xi)]}{\partial \tilde{\omega}}\big|_{\omega_0}$, and the ' symbol denotes differentiation with respect to frequency. The following scaling has been adopted: $\xi = z/\lambda_r$, $\tilde{y} = y/\lambda_r$, $\tau = ct/\lambda_r$, $\beta = 2\pi\tilde{\omega}$, and $\tilde{\omega} = \omega/\omega_r$, where $\lambda_r = 1\mu m$ is chosen as the reference wavelength.

Although Eqs.(4) contain no approximations, they may be simplified depending on the circumstances. In ordinary dielectric or semiconductor materials we may neglect second and higher order material dispersion terms, thus eliminating second and higher order temporal derivatives. As an example, in the spectral region of interest, which includes the entire visible and near IR ranges, the dielectric function (actual data) of $Si_3N_4$ [1] may be written as $\varepsilon(\tilde{\omega}) = 3.7798 + 0.17898\tilde{\omega} + 0.04408/\tilde{\omega}$. Using this approximately linear dielectric, one may estimate the second order dispersion length, defined as $L_D^{(2)} \sim \tau_p^2/|k''(\tilde{\omega})|$, where $\tau_p$ is incident pulse width, and $k''(\tilde{\omega}) = \partial^2 k/\partial \tilde{\omega}^2$. The result is $L_D^{(2)} \sim 2\times 10^3 \lambda_r$ (or ~2mm) for an incident, five wave-cycle pulse (~15fs); approximately 8mm for a ten wave-cycle pulse; and 1 meter for 100 wave cycle (~300fs) pulses.

In order to justify simplifications in Eqs.(4) when metals are present, one may proceed in a similar fashion. In the spectral region that includes a good portion of the visible range (~400-600nm), the real part of the dielectric function of silver, which follows a Drude model description [1], may be written as $\text{Re}[\varepsilon(\tilde{\omega})] = -3.54\tilde{\omega} + 31.1 - 65.1/\tilde{\omega}$. A similar expansion holds for the imaginary part. Thus, in a wavelength range about 200nm wide, the dielectric function of silver can easily accommodate a pulse only a few optical-cycles in duration, as long as propagation distances are smaller than typical material dispersion lengths. In our case, typical structure lengths are 500nm or less, far shorter than any typical, second order dispersion lengths associated with any metal, dielectric, or semiconductor material.

Another issue that is important for our purposes is the determination of energy and momentum in dispersive materials. We have recently addressed both issues [19,20], and here we provide the results. The energy density is derived from the general approach we have outlined above, and the result is [19]:

$$U(z,t) = \alpha_r |\mathcal{E}_x|^2 + \beta_r |\mathcal{H}_y|^2 + \frac{i\alpha_r'}{2}\left\{\mathcal{E}_x^* \frac{\partial \mathcal{E}_x}{\partial t} - \mathcal{E}_x \frac{\partial \mathcal{E}_x^*}{\partial t}\right\} + \frac{i\gamma_r'}{2}\left\{\mathcal{H}_y^* \frac{\partial \mathcal{H}_y}{\partial t} - \mathcal{H}_y \frac{\partial \mathcal{H}_y^*}{\partial t}\right\}$$
$$+ \frac{\alpha_r''}{6}\left(\mathcal{E}_x \frac{\partial^2 \mathcal{E}_x^*}{\partial t^2} + \mathcal{E}_x^* \frac{\partial^2 \mathcal{E}_x}{\partial t^2} - \frac{\partial \mathcal{E}_x}{\partial t}\frac{\partial \mathcal{E}_x^*}{\partial t}\right) + \frac{\gamma_r''}{6}\left(\mathcal{H}_y \frac{\partial^2 \mathcal{H}_y^*}{\partial t^2} + \mathcal{H}_y^* \frac{\partial^2 \mathcal{H}_y}{\partial t^2} - \frac{\partial \mathcal{H}_y}{\partial t}\frac{\partial \mathcal{H}_y^*}{\partial t}\right) + \ldots \quad (5)$$

where $\alpha_r = \text{Re}(\alpha), \gamma_r = \text{Re}(\gamma)$, and the ' symbol once again denotes differentiation with respect to frequency. The two leading terms correspond to the Landau energy density [21]. The additional terms are derived by interpreting the real and imaginary parts of the dielectric permittivity and magnetic permeability as being connected to energy flow and absorption, respectively. In reference [19] we showed that using this approach we are able to completely reconcile our results with Poynting's theorem:

$$\frac{\partial U}{\partial t} + Q_{dissipated} = \mathbf{E} \cdot \frac{\partial \mathbf{D}(\mathbf{r},t)}{\partial t} + \mathbf{H} \cdot \frac{\partial \mathbf{B}(\mathbf{r},t)}{\partial t} = -\nabla \cdot \mathbf{S}, \quad (6)$$

where

$$-Q_{dissipated}(z,t) = 2\omega_0 \left\{\varepsilon_i |\mathcal{E}_x|^2 + \mu_i |\mathcal{H}_y|^2\right\} + i\alpha_i \left(\mathcal{E}_x^* \frac{\partial \mathcal{E}_x}{\partial t} - \mathcal{E}_x \frac{\partial \mathcal{E}_x^*}{\partial t}\right)$$
$$+ i\gamma_i \left(\mathcal{H}_y^* \frac{\partial \mathcal{H}_y}{\partial t} - \mathcal{H}_y \frac{\partial \mathcal{H}_y^*}{\partial t}\right) - \frac{\alpha_i'}{2}\left(\mathcal{E}_x^* \frac{\partial^2 \mathcal{E}_x}{\partial t^2} + \mathcal{E}_x \frac{\partial^2 \mathcal{E}_x^*}{\partial t^2}\right) - \frac{\gamma_i'}{2}\left(\mathcal{H}_y^* \frac{\partial^2 \mathcal{H}_y}{\partial t^2} + \mathcal{H}_y \frac{\partial^2 \mathcal{H}_y^*}{\partial t^2}\right) + \ldots \quad (7)$$

is the rate at which energy is dissipated. Then we also show that for normal dispersion the Landau expression remains a good approximation to the energy density even in the limit of pulse durations that approach a single optical cycle, provided the dispersion functions vary relatively slowly with frequency, and that propagation distances are shorter than the second order dispersion length. These conditions are satisfied in our case to a high degree [14,19].

The other issue that should be considered is the form of the momentum density that we use. In a recent study [20] we found that, under conditions of material dispersion but negligible absorption, the conservation of linear momentum is consistent with the Lorentz force only if the Abraham electromagnetic momentum density, $\mathbf{g}_{\xi,\tilde{y}}(\tilde{y},\xi,\tau) = \mathbf{S}_{\xi,\tilde{y}}(\tilde{y},\xi,\tau)/c^2$, is used [24], where

$$\mathbf{S} = \frac{c}{4\pi}\mathbf{E} \times \mathbf{H} = \hat{\mathbf{y}}S_{\tilde{y}}(\tilde{y},\xi,\tau) + \hat{\mathbf{z}}S_\xi(\tilde{y},\xi,\tau) = \frac{c}{4\pi}\left(\hat{\mathbf{y}}(\mathcal{H}_x \mathcal{E}_\xi^* + \mathcal{H}_x^* \mathcal{E}_\xi) - \hat{\mathbf{z}}(\mathcal{H}_x \mathcal{E}_{\tilde{y}}^* + \mathcal{H}_x^* \mathcal{E}_{\tilde{y}})\right) \quad (8)$$

is the Poynting vector. Based on the results of reference [20], we are thus well-justified in choosing the Abraham momentum density, even when absorption is no longer negligible, because it provides the basis for momentum conservation in the absence of absorption.

**Energy and Momentum Distribution in Transparent Metal Stacks**

The total, instantaneous momentum for a wave packet of finite spatial extension is defined as follows:

$$\mathbf{P}(\tau) = P_{\tilde{y}}(\tau)\hat{\mathbf{y}} + P_{\xi}(\tau)\hat{\mathbf{z}} = \hat{\mathbf{y}} \int_{\xi=-\infty}^{\xi=\infty} \int_{\tilde{y}=-\infty}^{\tilde{y}=\infty} g_{\tilde{y}}(\tilde{y},\xi,\tau)\, d\tilde{y}\, d\xi + \hat{\mathbf{z}} \int_{\xi=-\infty}^{\xi=\infty} \int_{\tilde{y}=-\infty}^{\tilde{y}=\infty} g_{\xi}(\tilde{y},\xi,\tau)\, d\tilde{y}\, d\xi \quad (9).$$

We may also define the Snell or refraction (phase front) angle as the expectation value of the vector sum of the individual wave vectors along the two spatial coordinates:

$$<\mathbf{k}(\tau)> = \int_{k_{\xi}=-\infty}^{k_{\xi}=\infty} \int_{k_{\tilde{y}}=-\infty}^{k_{\tilde{y}}=\infty} \left(\hat{\mathbf{y}}k_{\tilde{y}} + \hat{\mathbf{z}}k_{\xi}\right) |H(k_{\tilde{y}},k_{\xi},\tau)|^2 \, dk_{\tilde{y}} dk_{\xi} \quad . \tag{10}$$

Although for individual plane waves the direction of the Poynting vector coincides with the direction of the electromagnetic momentum density, for localized wave packets it is always more appropriate to talk about momentum. One then generally finds that, as defined in Eqs.(9-10), **P** and <**k**> are co-directional in ordinary, isotropic, low-absorption materials, and that in negative index materials they point in opposite directions [18], as expected [3].

Eqs.(9-10) represent instantaneous quantities. Taken together they may be used to calculate the refraction angles of momentum (or energy) and phase front. As mentioned earlier, we use the word "refraction" with some caution, because it is usually intended to mean the direction of the wave vector **k**. However, this parallelism ceases to hold in metals, for both TE and TM polarized incident waves. The consequences are more dramatic in the case of TM polarized fields, because they lead to anomalous refraction of the Poynting vector, as discussed in reference [2], even though waves are generally evanescent. We should point out that the expression "evanescent waves" is typically used to describe a steady state condition. Full knowledge of the dynamical aspects of the interaction can be gained only by solving the equations of motion in the time domain in the presence of dispersion. This then reveals that energy and momentum generally reside with the fields, regardless of their location or the nature of the material. The imaginary part of the index of refraction, which is large in metals, is simply an indication that field oscillations occur under strongly damped conditions. But this effect does not by itself invalidate Poynting's theorem, whose validity continues to hold because it is a statement about energy conservation. In other words, one can and should refer to energy and momentum as refracting inside any absorbing material, including metals. The angle of refraction may be complex in general, but a direction of motion for momentum and energy can still be defined and identified, as long as no contradictions arise. This phenomenon clearly manifests itself inside a typical transparent metal stack, as energy and momentum become localized inside the metal layers in equal or greater amounts compared to the energy and momentum found inside the dielectric layers. This is the process that we describe below.

To illustrate the negative refraction process through a transparent metal stack, we examine the propagation of short pulses through a symmetric 6 and 1/2 period, Ag(32nm)/X(21nm) stack. Here, X represents a generic, high-index material that we use to show how the process works, and at the end we will provide a list of materials suitable for the visible range. We assume the index of refraction of material X is n=4. The stack thus contains 192nm of silver, and it is approximately 318nm thick. Therefore, about 60% of the structure is silver, and it is difficult to justify any kind of effective medium approach to describe this type of structure. The first and last layers of the stack are X layers, each approximately 11nm thick. This choice is useful because it tends to smooth and maximize the transmission function, and to minimize reflections [14].

In Fig.1 we depict the plane-wave transmittance through the stack as a function of wavelength, for TM-polarized light incident at 45°. We tune the carrier wavelength of the incident pulse at 400nm, where the transmittance is ~30%. The dielectric constant of Ag at 400nm is $\varepsilon = -3.77 + i0.67$ [1]. We note that we use actual material data already tested and verified specifically for transparent metal stacks [10-13]. For simplicity, but only for the moment, we assume that material X is dispersion-less and absorption-less. We assume our incident pulse is Gaussian, and ~30fs in duration (1/e width), with carrier frequency tuned to 400nm. At this wavelength, the pulse is ~20 optical cycles in duration, so that the curvature of the wave fronts changes little over the entire propagation distance. Fig.(2) contains several snapshots that chronicle the dynamics of the pulse as it traverses the stack. The centroid of the transmitted wave packet appears to be shifted upward, a detour that is best perceived for short, well-localized wave packets, but that clearly holds for any pulse of larger but finite spatial extension and longer duration. The movie associated with Fig.(2) follows the pulse as it interacts with the metal-dielectric stack. This shift can only be understood in terms of additional upward momentum supplied to the pulse by the structure.

Fig.(3) summarizes the negative refraction process and our basic result. Although the quantities we calculate are instantaneous, one may proceed by either performing a time average as the pulse dwells inside the stack, or by simply taking the most dominant values, which occur when the peak of the pulse arrives at the structure in order to mimic near plane-wave conditions. The total momentum refracts upward inside the transparent metal stack, forming an angle of approximately -40º with respect to the normal. Given a structure length of ~318nm, the upward shift, denoted by $\delta$ in the figure, is given by $\delta \sim 266$nm. Furthermore, calculations similar to those exemplified in Fig.(2) were carried out for incident angles of 30º and 60º. The resulting *momentum-energy* refraction angles were found to be ~ -27º and ~ -51º, respectively. Application of Snell's law reveals that all these angles have in common an *effective*, negative refractive index of $n \sim -1.1$. These propagation exercises thus reveal that a strongly diffracting wave containing a span of transverse k-vectors, emanating from a small aperture and traversing the transparent metal stack, will be focused at a point because the effective index of refraction is the same for all angles of incidence investigated. However, this result is rendered somewhat dramatic by the fact that the total momentum and the total wave-vector *do not* point in the same direction, and in fact in this case they point in altogether different quadrants. Yet, Snell's law still appears to work. Of course, one cannot ascribe to the wave vector its usual meaning and qualities, and a clear inequality between group and energy velocity quickly emerges, with the group velocity becoming an ill-defined concept.

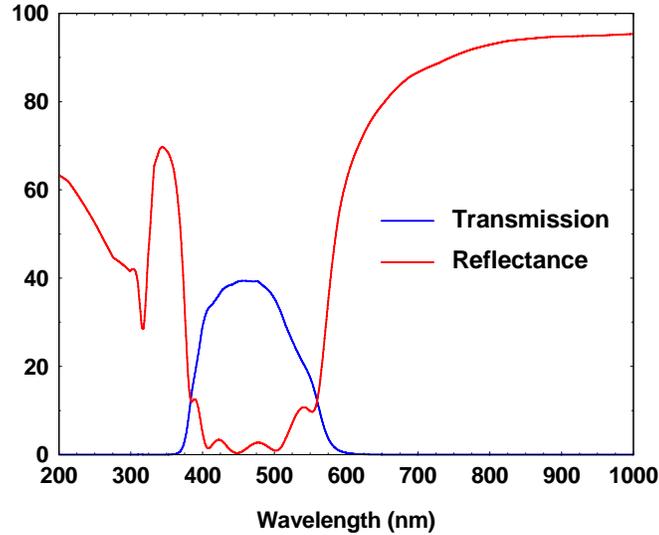

**Fig.1**: Transmittance vs. wavelength from a 6 and 1/2 period stack composed of Ag(32nm)/X(21nm), inclusive of entry and exit X layers 11nm thick. The angle of incidence is 45º. The transmission window extends over a good portion of the visible range.

Although this result may seem surprising, the fact is that it can be shown that energy and group velocities are the same only for homogenous, absorption-less media, away from any boundaries [22]. It has been demonstrated that in the absence of absorption the two velocities are generally different near surfaces or inside strongly scattering one-dimensional systems, such as photonic band gap structures of finite size [23]. We do not associate this discrepancy with anisotropic behavior because it can be understood in simpler terms, having to do with pulse reshaping and scattering losses due to geometrical dispersion, i.e. the introduction of longitudinal index discontinuities. Nevertheless, the physical difference between group and energy velocity is generally not recognized outside the realm of anisotropic materials, even in structures of finite length, and in most of the literature the terms are used interchangeably. Then, it should not be surprising to discover that in absorbing, two-dimensional systems such as ours energy and group velocities cease also to be collinear, and not easily related in an obvious manner. In addition, a study of the Fourier spectrum of the pulse, similar to that done in reference [18], clearly shows that during the entire interaction the spectral content of the pulse does not display any components that point in the same quadrant as the momentum vector.

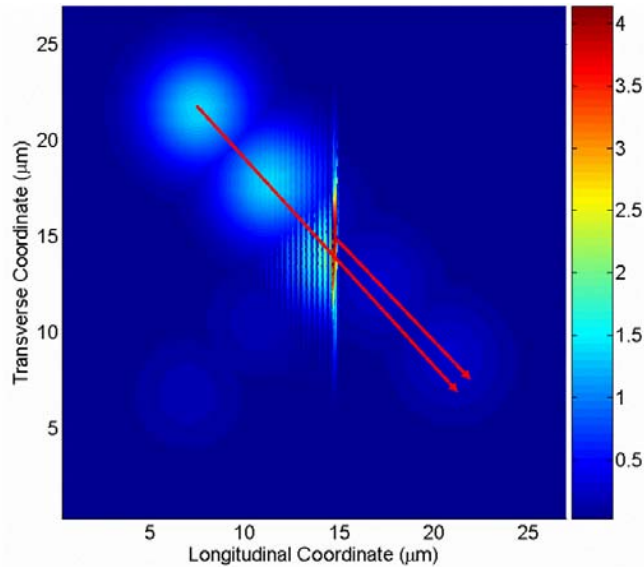

**Fig.2**: An incident Gaussian, TM-polarized wave packet is incident on the transparent metal described by picture showing several snapshots of the pulse in Figure 1. The centroid of the pulse that exits to the right of the stack is shifted upward by approximately 266nm.

Now that we have established that negative refraction occurs in the visible range, we wish to test the performance of this particular structure across the transparency window. We fix the incident angle at 45º, tune the carrier frequency of the pulse at 500nm, and find a negative refraction angle of ~ -7º. We then tune the pulse at 600nm, and find that the pulse still refracts negatively, but with an angle of ~ -2º. Although the basic cause and effect for the pattern that emerges is quickly identified, namely that the magnitude of the dielectric constant of silver eventually overtakes that of the dielectric material ($\varepsilon$=16) at longer wavelengths, more is at play here than just the ratio of dielectric constants. At 400nm, the magnitude of the ratio between the real parts of the dielectric constant is 16/3.77. At 500nm the ratio drops to 16/8.57, and by the time we reach 600nm the ratio is ~16/14. With reference to the effective medium approach set forth in reference [17], one may be tempted to extrapolate those results to the present stack. That is, as the condition of equal and opposite dielectric constants is reached, the refraction angle quickly approaches zero. However, this view masks the physical mechanisms that are actually responsible for the behavior. The refraction process is generally always anomalous (upper, right quadrant) in each metal layer, and normal (lower, right quadrant) in each dielectric layer. This aspect of the dynamics is exemplified on the right, lower side of Fig.(3), where the arrows depict the momentum vectors in each type of layer. The high index of refraction in the dielectric layer leads to smaller refraction angles inside that layer. At the same time, and more importantly, the electric and magnetic fields become localized inside the metal layers in almost equal measure compared to the dielectric layers [10-16].

There is an issue related to the nature of the fields inside the stack that merits some elucidation, in order to connect to, and emphasize, previous comments that we made. When one thinks of fields' distribution inside thick, isolated metal films, such as those studied in reference [2], there is a tendency to speak of evanescent waves. Earlier we noted that in the visible range, the imaginary part of the index of refraction can be more than one order of magnitude larger than the real part, causing the portion of the field that enters the metal to be oscillatory, but strongly damped. What we are describing here is a structure that contains individually thick metal layers such that energy and momentum (at least according to the classical description presented in the previous sections) are almost equally distributed inside both the metal and the dielectric material. In the case of negative refraction, the metal is able stack described in Figure 1, for a pulse at normal incidence. The real part of the dielectric constant alternates between the values of 16 and -3.77, and its value is plotted on the right axis. We note that both fields are intense inside the each metal layer, a fact that leads to large energy and momentum values inside the metal.

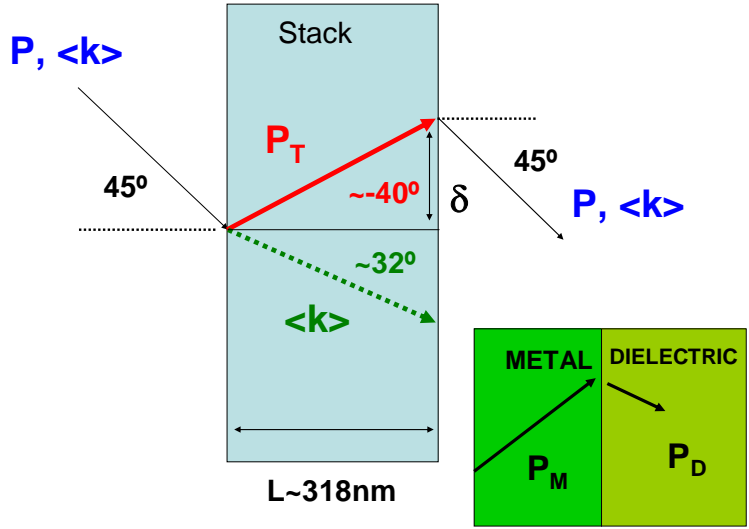

**Fig.3**: Schematic representation of the refraction that occurs inside the stack (left drawing). Although **P** and **k** point in different directions, the angle of refraction is still predicted by Snell's law. Lower right drawing: the local momenta inside two adjacent metal and dielectric layers are shown.

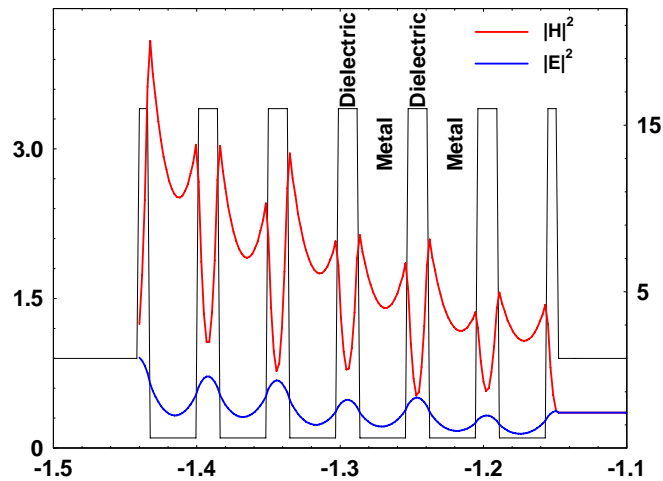

**Fig.4**: On-axis electric and magnetic field intensities vs. position inside the multilayer

to accumulate and store more energy and momentum than the dielectric material thanks to the resonance tunneling conditions allowed by an appropriate choice of dielectric materials and thicknesses. In Fig.(4) we show the electric and magnetic field intensity distributions for the case of normal incidence on the stack in question. It is clear that magnetic and electric fields become simultaneously localized inside the metal layers, thus yielding an unusually large momentum and energy concentration, which can be exploited in various ways, for instance, to access the large $\chi^{(3)}$ of copper [16]. However, it should also be understood that the fact the light makes its ingress in metal layers, thanks to the resonance tunneling condition, does not automatically translate into large absorption. This is also an experimentally verified fact [11,12,13,16].

In bulk metals, or single, thick, isolated metal layers, the electric and magnetic fields are out of phase by approximately 45º [24], as the electric field is expelled from the metal. This phase mismatch normally translates into a spatial delocalization of the fields, and the fields are strongly damped. In sharp contrast, inside our transparent metal stack this spatial delocalization does not hold, as both electric and magnetic fields simultaneously occupy each metal layer, as seen in Fig.(4). As a result, and in addition to pointing in the anomalous direction, the local Poynting vector (and resulting momentum), which is generally the product of the electric and magnetic fields, can be large inside the metal. For the case in question, the total momentum inside the metal layers displays an angle of approximately 58º, larger than it is possible to achieve inside any single metal layer [2], at least in a high transmission regime. If this is combined with the fact that the dielectric momentum makes a shallower angle with the normal (Fig.(3)), yielding smaller downward components, then the metal-momentum dominates and yields a total momentum that points in the anomalous direction. These dynamics are completely determined by an interference phenomenon that generally requires relatively thick metal layers, unlike the approach outlined in reference [17], where the field is described as a mean field, where no interference effects are recorded, and where no negative refraction takes place.

**Sub-wavelength Resolution**

Up to this point we have only addressed the issue of the existence of negative refraction. However, we also wish to tackle the subject of super-lensing, and whether it is possible to resolve an object below the wavelength limit. We begin by once again considering a pulse tuned at 400nm, but this time it is incident on an opaque silver shield with a small aperture. The shield is approximately 200nm thick, opaque by any measure, and the slit is ~125nm wide. The situation is shown in Fig.(5). The slit, which has depth and resembles a waveguide, is located ~50nm away from the stack. The image plane is located approximately 570nm away from the slit, or ~200 nm away from the stack. We collect the image at this location because a ray-tracing diagram using the refraction parameters distilled form Fig.(2) suggests the external focus should be found near that point. Fig.(6) captures a snapshot of the field, just as part of the wave packet forms a clear focus on the right side of the stack, at an approximate distance of 200nm. Field localization and hot spots that accompany the internal focus, which appears to form closer to the exit side, are visible inside the multilayer stack. In the figure we are plotting the magnetic field intensity, which according to Fig.(4) is more intense within the metal layers. The approximate center of the external focus is intersected by the imaging plane (thin yellow line), where we sample the field. The resulting transverse field profile is a typical diffraction pattern, and is shown on Fig.(7). The slit casts a bright, discernable spot, with FWHM ~200nm wide, it is resolved ~584nm away, with an incident wavelength roughly four times larger than slit size. In comparison, the transverse profile of a field diffracting in free space from a similar slit is approximately ten times broader at the same location. In Fig.(8) we depict the interference pattern that results when the pulse is incident on two small, adjacent apertures. The slits are 125nm wide, and they are separated by an opaque region ~175nm wide, so that their center-to-center distance is ~300nm.

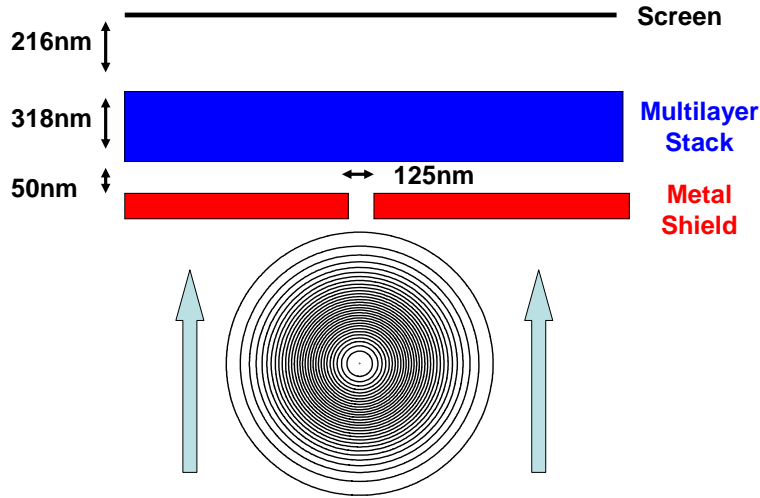

**Fig.5**: A Gaussian wave packet is incident form the bottom on a metal shield with an aperture ~125nm wide. The transparent metal stack described in the caption of Fig.1 is located ~50nm away from the slit. An image plane stands at ~584nm from the aperture.

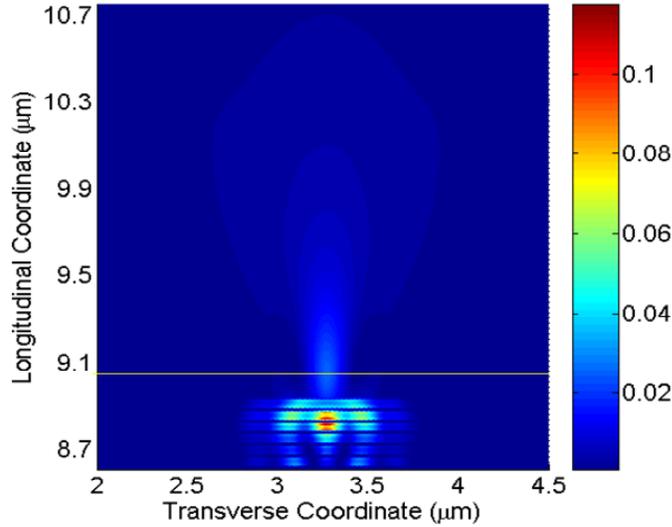

**Fig.6**: Contour diagram of a snapshot of the field inside and above the transparent metal stack. The figure shows that focal points are visible inside and just above the stack, about 216nm away from it. The yellow line intersects the focus, and represents the position of an imaging plane.

In Fig.(9) we image the field approximately 100nm down-range from the stack, and find that the two slits can be resolved with a visibility of approximately 50%. Although the conditions that we have studied are not optimized, we already see clear evidence of sub-wavelength resolution. Under the conditions we have described, each aperture is thus able to form its own spot. However, in Fig.(8) a third spot is also visible down-range: the two spots become secondary sources and trigger the onset of a classic

Poisson spot phenomenon. We have simulated situations with four slits, and it becomes increasingly clear that the resulting interference grid resembles a checkerboard pattern, with multiple layers of interference and Poisson spots that carry the light forward.

A comment about the role that the metal screen plays is in order. One may be tempted to think that plasmonic activity at the aperture on the metal screen, is at least partially responsible [25] for the focusing mechanism that we record inside the stack, and then again on the other side of it. In fact, our simulations suggest that replacement of the silver shield with a shield that does not support plasmonic activity at the carrier wavelength, such as tungsten, or a completely absorbing dielectric layer with a slit, leads to precisely the same results. Therefore, we can confidently state that plasmonic activity at or near the aperture of the silver shield plays no role in the results that we report here.

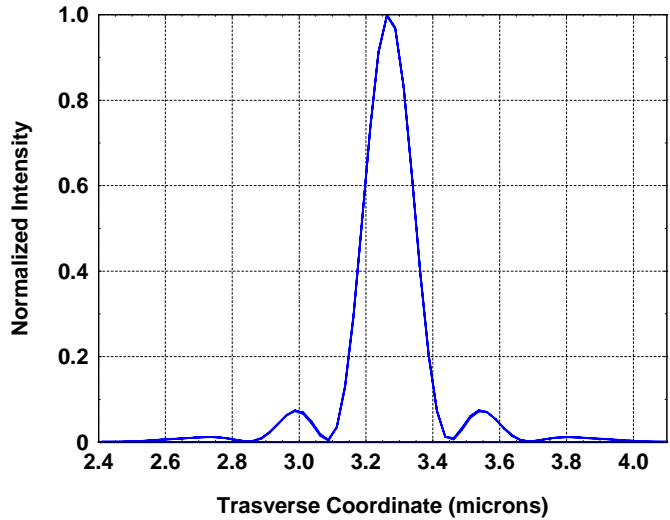

**Fig.7**: Image produced by the slit on the image plane indicated on Figure 6. The full width at half maximum of the field that propagates through the stack (solid line) is ~200nm and it is approximately ten times narrower than the field emanating from the same aperture, and propagating in free space.

We have identified several candidates that are suitable in the visible range, for example Si, GaP, and GaAs. The indices of refraction of Si and GaAs are above 4, while that of GaP is between 3 and 4. Although bulk Si and GaAs are opaque in the visible range, a number of thin layers may be combined with varying thicknesses of Ag to yield a transparency range, as a simple, standard, matrix-transfer calculation quickly reveals. For instance, we find that 6 and 1/2 periods of Ag(32nm)/Si(22nm) yield a transmittance window centered around 500nm (where the real part of the dielectric constant of Si is ~18), with maximum transmittance of ~20%. At this wavelength, our model yields a negative refraction angle of ~ - 10º. Even though GaP has a lower index of refraction than Si, one may obtain substantially similar results with a stack that contains almost 260nm of Ag. Finally, at 632nm we find that the transmittance from a 6 and 1/2 period Cu(32nm)/Si(43nm) is still approximately 30%, and yields a negative refraction angle of ~ -3º. So, it is possible to find common materials and design transparent metal stacks, and still have negative refraction ad sub-wavelength resolution occur in the visible range.

**Conclusions**

In conclusion, we have presented theoretical evidence that negative refraction can occur throughout the visible range in transparent, metallo-dielectric, photonic band gap structures. These structures may contain hundreds of nanometers of metal, and yet they can be transparent to visible light. Our simulations show that incident TM-polarized waves undergo negative refraction, and that the stack may form an image resolved below the wavelength limit, with possible further optimizations and improvements. These results open a new point of view regarding the propagation of light in metals, and may help pave the way to the achievement of practical devices and applications based on negative refraction, with realistic numerical apertures. Since transparent metal structures have already been fabricated in the visible spectrum using a number of different dielectric materials, the experimental realization of negative refraction in such structures presents no new discernable technical challenges. Taken together with the tunable aspects of the process, which we have touched upon, the search for suitable, high-index materials may yield candidates that could also make it possible to increase the object-lens distance, and to push the process well into the infrared region of the spectrum, with possible, significant applications to IR imaging.

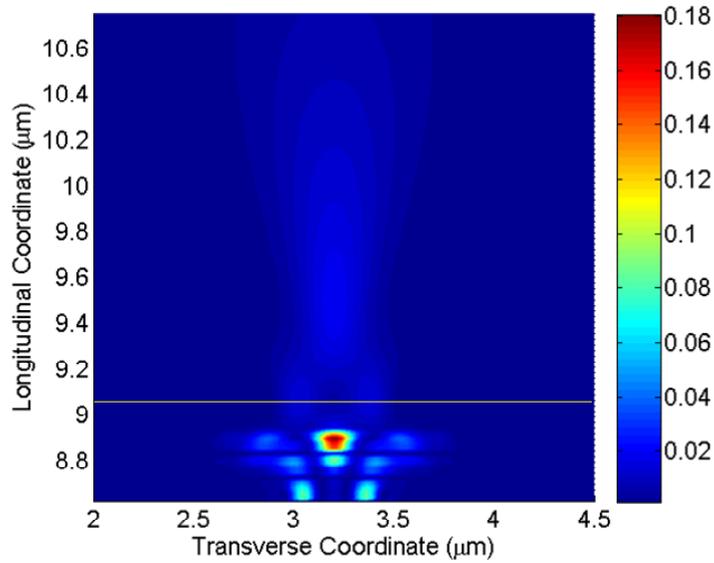

**Fig.8**: Contour diagram of a snapshot of the field inside and above the transparent metal stack. In this case the image is produced by two 125nm apertures located on the metal shield, having a center-to-center distance of 300nm. The figure shows that each slit forms its own image. A third bright spot appears down-range, as a result of constructive interference between the primary spots, which become secondary sources.

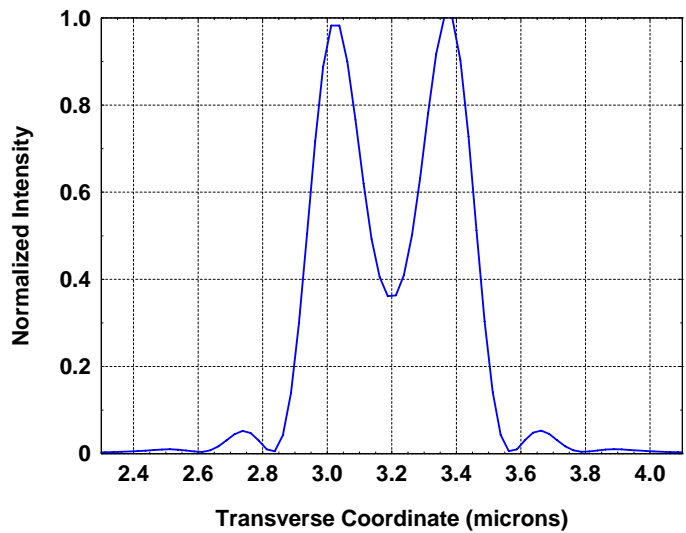

**Fig.9**: Image produced by the two-slit system described in the caption of Fig.(8), a distance of ~100nm from the multilayer stack. The figure suggests a fringe visibility better than 50%.